\documentclass{KapProc} 

\usepackage{t1enc}
\usepackage{ProcPs} 

\let\footnote\savefootnote
\let\footnotetext\savefootnotetext 
\let\footnoterule\savefootnoterule 

\kluwerbib

\startauthorindex

\usepackage{epsf}
\def\ga{\mathrel{\mathchoice {\vcenter{\offinterlineskip\halign{\hfil
$\displaystyle##$\hfil\cr>\cr\sim\cr}}}
{\vcenter{\offinterlineskip\halign{\hfil$\textstyle##$\hfil\cr  
>\cr\sim\cr}}}
{\vcenter{\offinterlineskip\halign{\hfil$\scriptstyle##$\hfil\cr
>\cr\sim\cr}}}  
{\vcenter{\offinterlineskip\halign{\hfil$\scriptscriptstyle##$\hfil\cr
>\cr\sim\cr}}}}}
\begin{document}

\articletitle[AGB and Post-AGB Evolution: Structural and Chemical Changes]
{AGB and Post-AGB Evolution: \\ Structural and Chemical Changes
}

\author{T.\ Bl{\"o}cker, R.\ Osterbart, G.\ Weigelt}
\affil{Max-Planck-Institut f{\"u}r Radioastronomie,
        53121 Bonn, Germany}
\author{Y.\ Balega}
\affil{Special Astrophysical Observatory,
       Nizhnij Arkhyz,
       357147, Russia}
\author{A.\ Men'shchikov}
\affil{Stockholm Observatory, 13336 Saltsj{\"o}baden, Sweden}



\begin{abstract}
Structural and chemical changes during the AGB and post-AGB
evolution are discussed with respect to
two recent observational and theoretical findings.
On the one hand, high-resolution infrared observations 
revealed details of the 
dynamical evolution of the fragmented, bipolar dust shell
around the far-evolved carbon star IRC\,+10\,216 
giving evidence for rapid changes of an already  PPN-like structure during
the very end of the AGB evolution. 
On the other hand, stellar evolution calculations considering convective
overshoot have shown how thermal pulses  during the post-AGB stage lead to the
formation of hydrogen-deficient post-AGB stars with abundance patterns 
consistent with those observed for Wolf-Rayet central stars.
\end{abstract}

\section{Introduction}
During the evolution along the Asymptotic Giant Branch (AGB) and the
subsequent post-AGB phases various structural and chemical changes of the
stars and their surrounding dust-shells and nebulae take place.
One important issue is  the change of the dust-shell symmetry
during transition from the AGB to the post-AGB stage.
Unlike dust-shells around AGB stars, post-AGB objects as
protoplanetary nebulae often expose prominent features of
asphericities, in particular in axisymmetric geometry (e.g.\ Olofsson 1996),
The shaping of planetary nebulae can be described by
interacting stellar wind theories (see Kwok 2000)
treating the interaction of a fast (spherical) central-star wind
with the preceding slow (aspherical) AGB wind. 
The establishment of bipolar structures
seems to  begin
already during the (very end of) AGB evolution.
One example of this stage of evolution is the carbon star \inx{IRC\,+10\,216}.
Its recent high-resolution observations
indicate asymmetric mass-loss processes and even give insight into its
dynamical dust-shell evolution (Osterbart et al.\ 2000).

Another matter of debate is the 
origin of hydrogen-deficient post-AGB stars. Although stars
evolving through the AGB phase stay hydrogen-rich at their
surfaces, a considerable fraction of their descendants, the
central stars of planetary nebulae (CSPNe),
show hydrogen-deficient compositions (Mendez 1991).
Approximately 20\% of the whole CSPNe population
appears to be hydrogen deficient
while the rest show solar-like compositions.
Important constituents of the hydrogen-deficient population
are the Wolf-Rayet (WR) central stars and the hot
PG\,1159 stars with typical surface abundances of (He,C,O)=(33,50,17)
by mass (Dreizler \& Heber 1998, Koesterke \& Hamann 1997).
Standard stellar evolution calculations failed to model these objects
since they predict post-AGB stars either to have hydrogen-rich surfaces 
(e.g.\ Bl\"ocker 1995) or, if hydrogen-deficient,
to expose only a few percent of oxygen in their photospheres
(Iben \& McDonald 1995).
However, if convective overshoot is considered,
hydrogen-deficient post-AGB stars with
abundance patterns as observed in WR central stars can be formed
(Bl\"ocker 2000, Herwig 2000).

\section{The dust-shell of \protect\inx{IRC\,+10\,216}}

\inx{IRC\,+10\,216}\ (\inx{CW Leo})
is the nearest ($d\sim 130$\,pc, Groenewegen 1997)
and best-studied carbon star and
one of the brightest infrared sources in the sky. 
Due to strong stellar winds of $\dot{M}
\approx 2-5\times10^{-5}$M$_{\odot}\,$yr$^{-1}$ (Loup et~al.\ 1993)
it is highly enshrouded by dust.
The central star of IRC\,+10\,216 is a
long-period variable with a period of $\sim 649$\,days (Le~Bertre 1992).
The bipolar appearance of the dust shell 
around this object was reported by Kastner \& Weintraub (1994).
The non-spherical structure is
consistent with the conjecture that IRC\,+10\,216 is in a phase
immediately before entering the protoplanetary nebula stage. 
High-resolution observations showed that the 
inner circumstellar dust shell is fragmented
(Weigelt et~al. 1998, Haniff \& Buscher 1998).
The results of Dyck et~al.\ (1991) and Haniff \& Buscher (1998)
showed that the dust-shell structure of IRC\,+10\,216 is changing
within a timescale of only several years. The dynamical dust-shell
evolution was revealed in detail by recent observations
of Osterbart et al.\ (2000) and Tuthill et al.\ (2000).

\begin{figure}
  \setlength{\unitlength}{0.5mm}
  \begin{picture}(146,365)(0,0)
    \put(0,292){ \epsfxsize=36mm\epsfbox{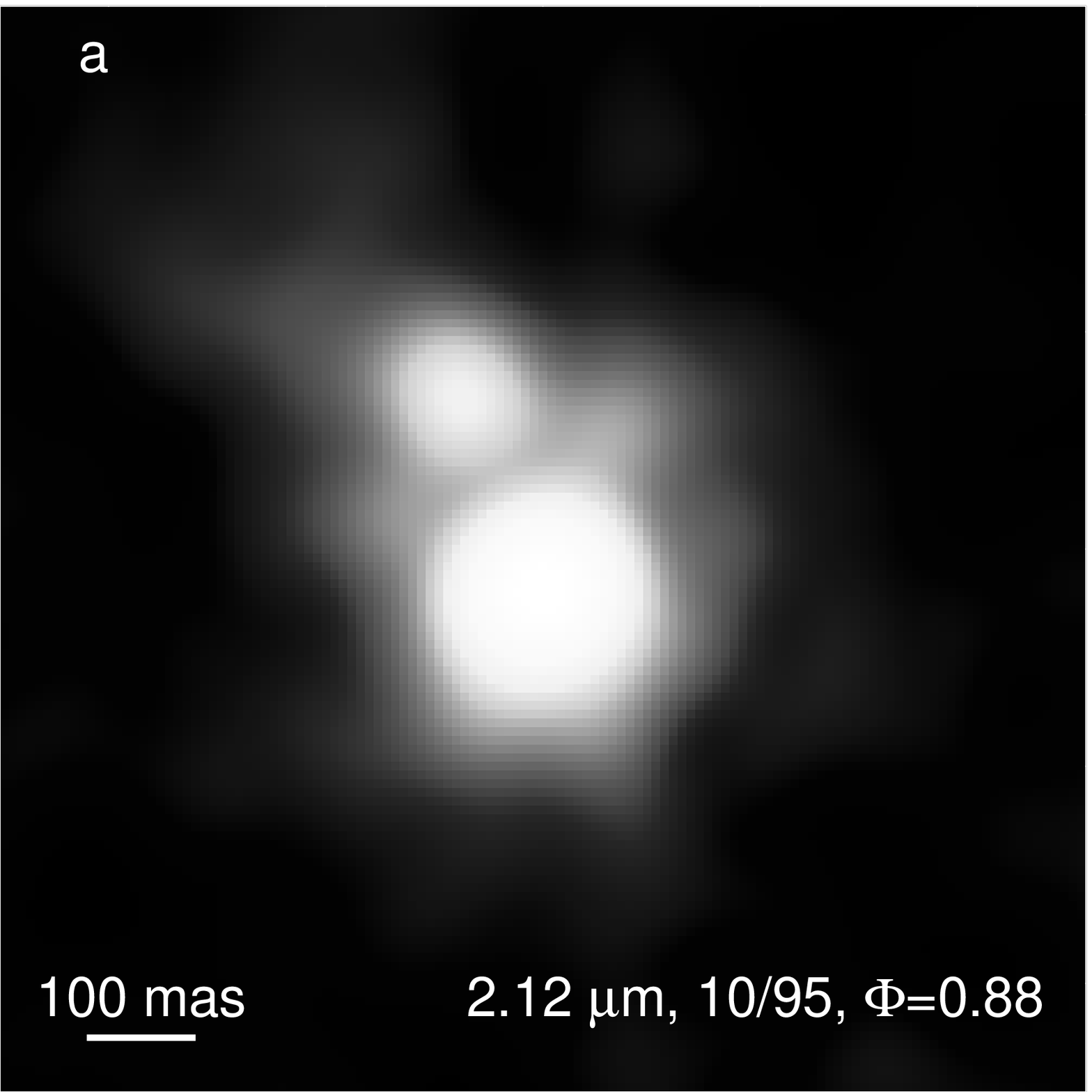}    }
    \put(0,219){ \epsfxsize=36mm\epsfbox{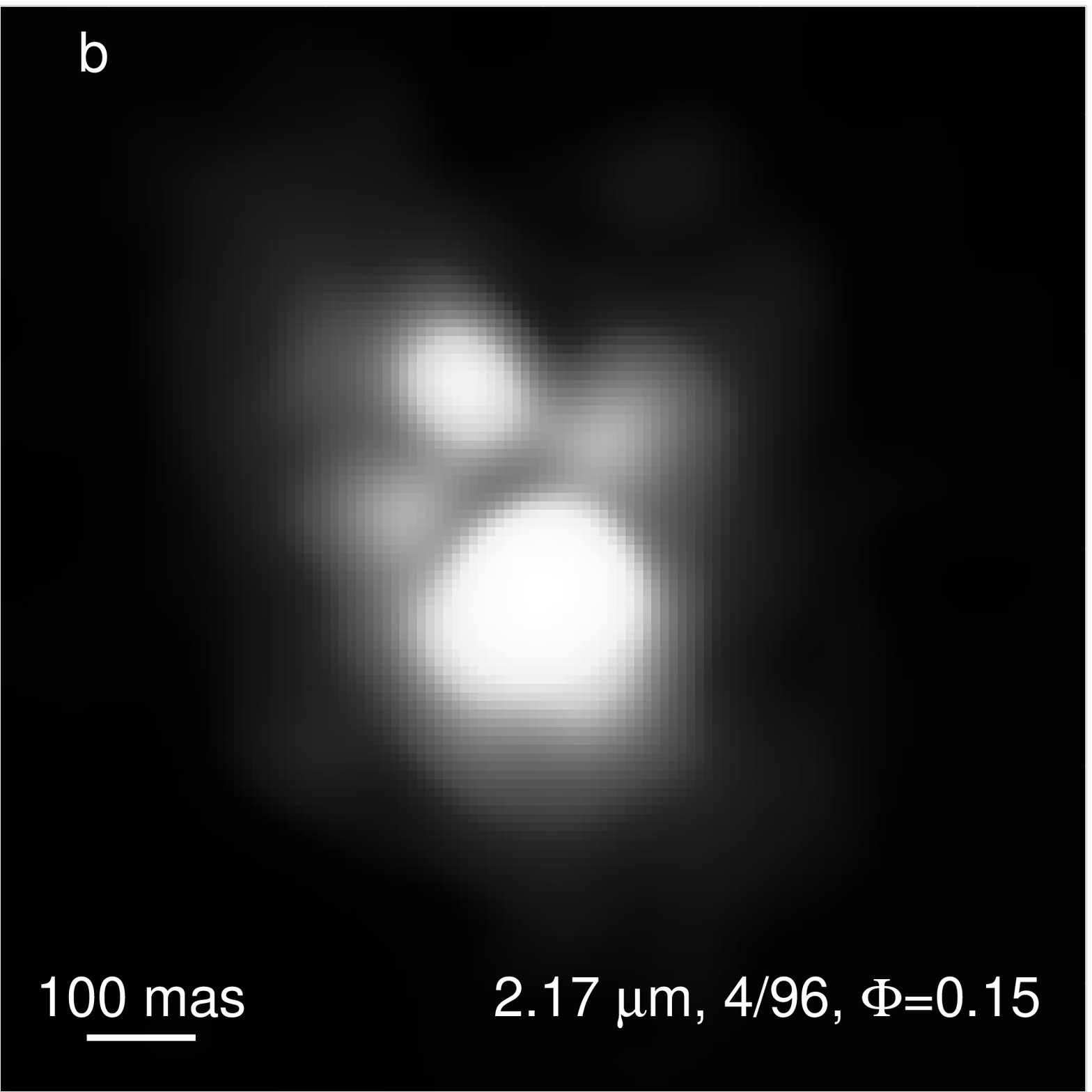}    }
    \put(0,146){ \epsfxsize=36mm\epsfbox{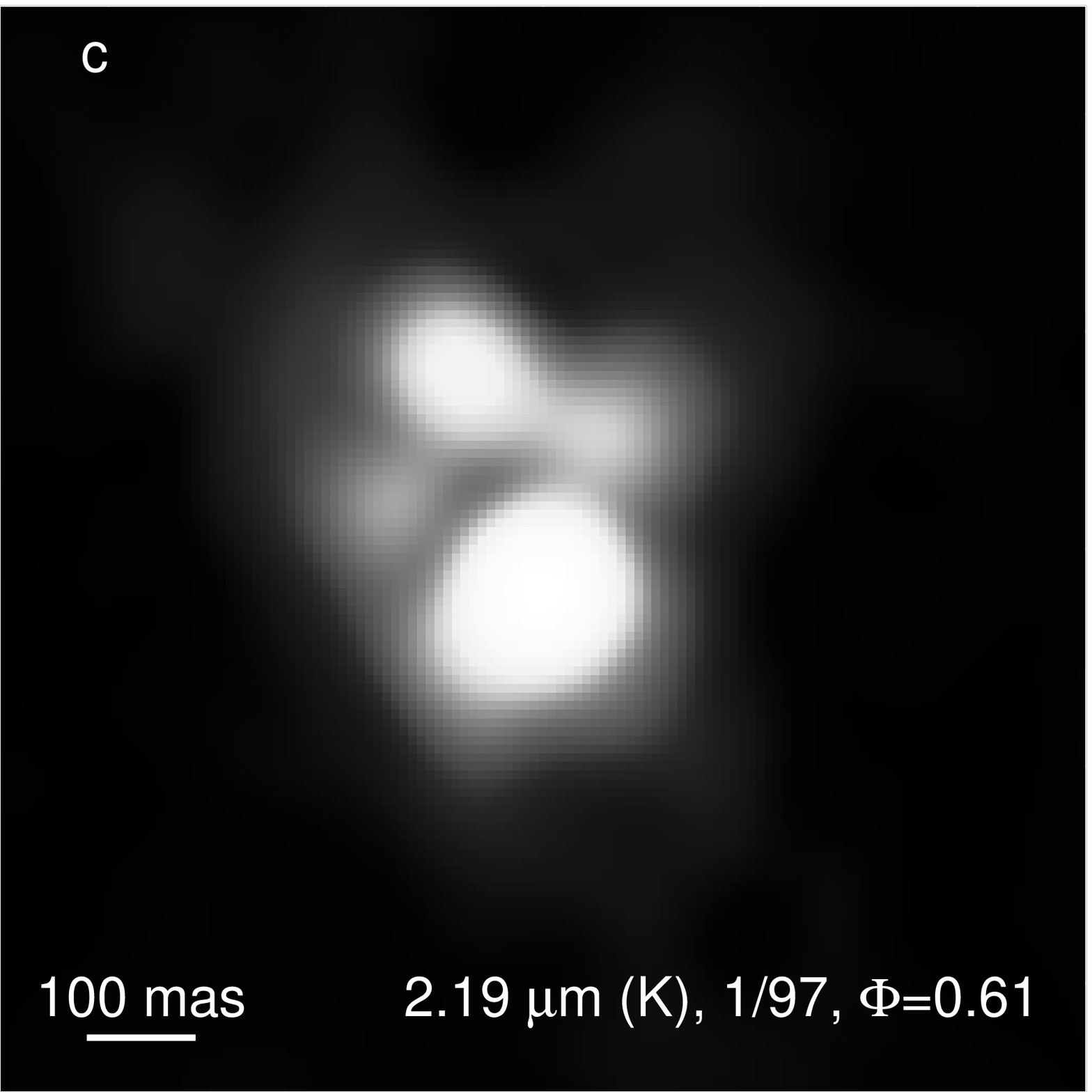}     }
    \put(0,73){  \epsfxsize=36mm\epsfbox{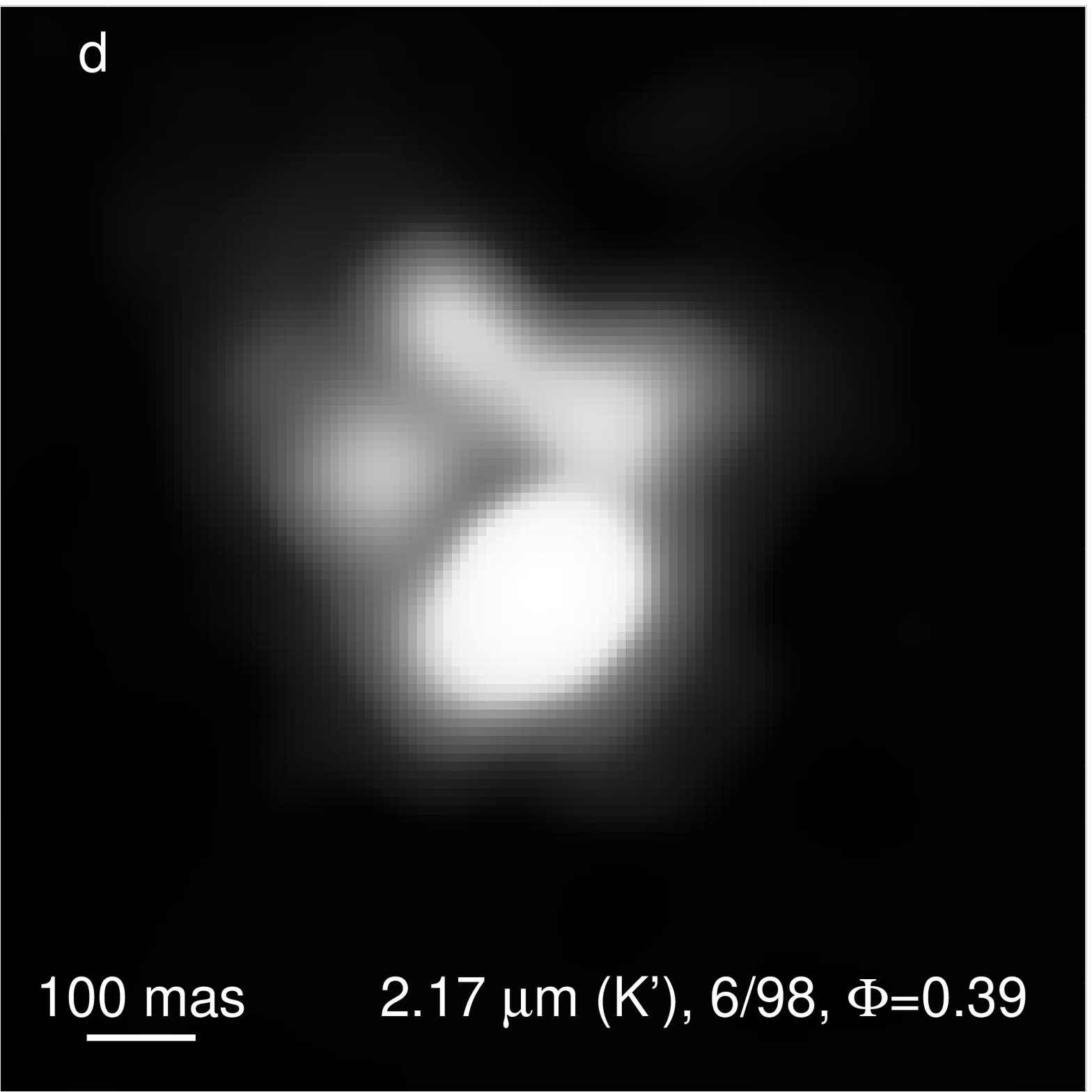}    }
    \put(0,0){   \epsfxsize=36mm\epsfbox{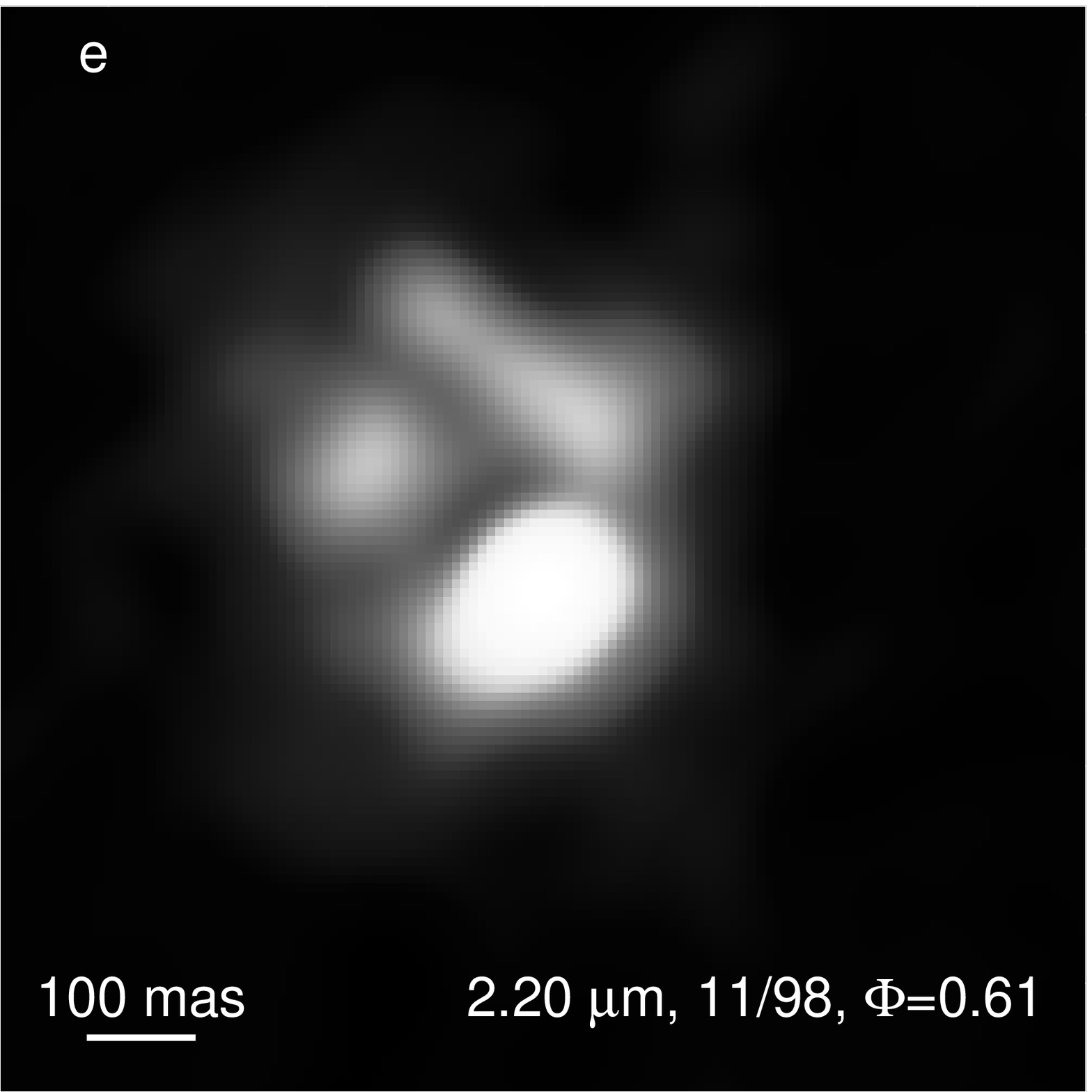}   }
    \put(73,292){ \epsfxsize=36mm\epsfbox{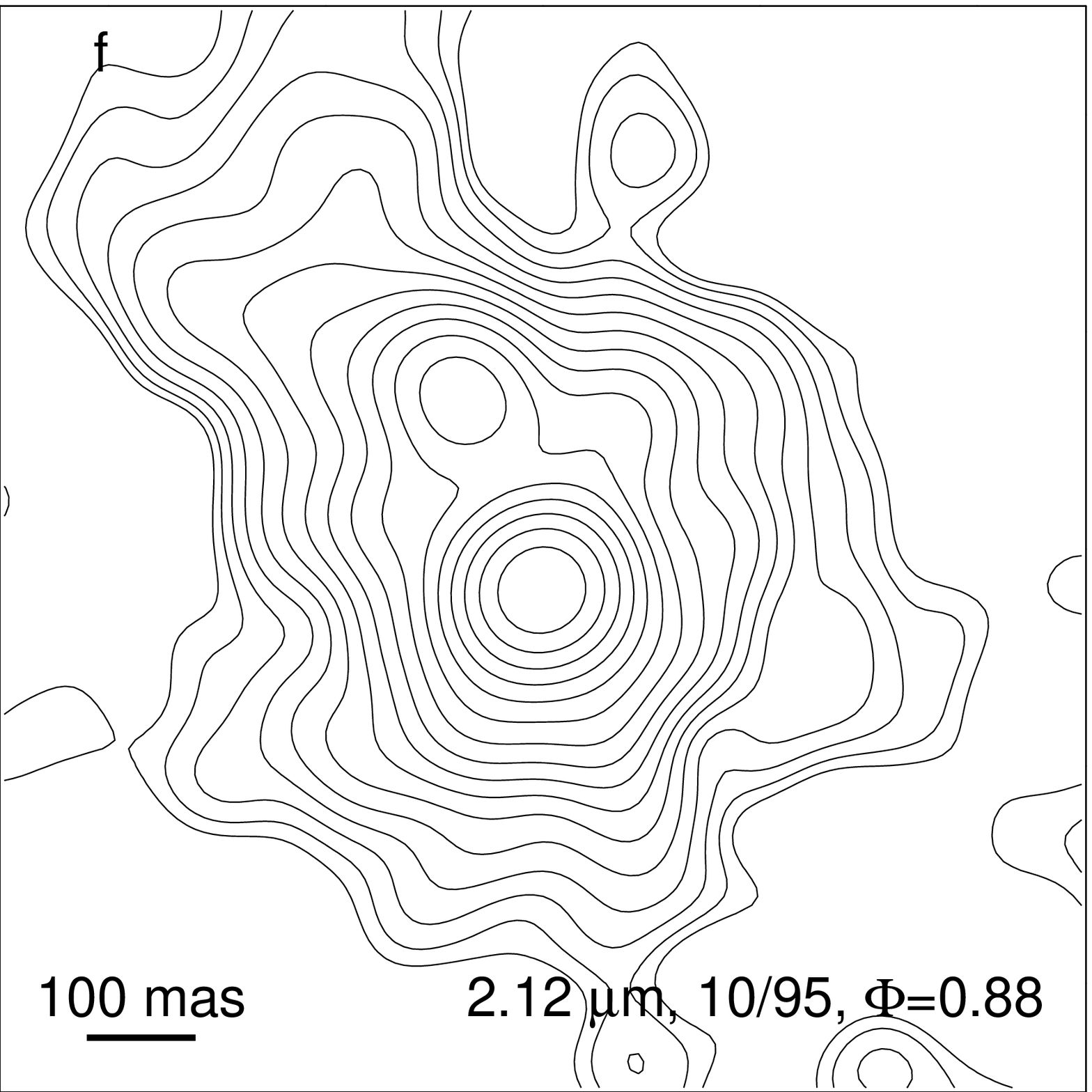}  }
    \put(73,219){ \epsfxsize=36mm\epsfbox{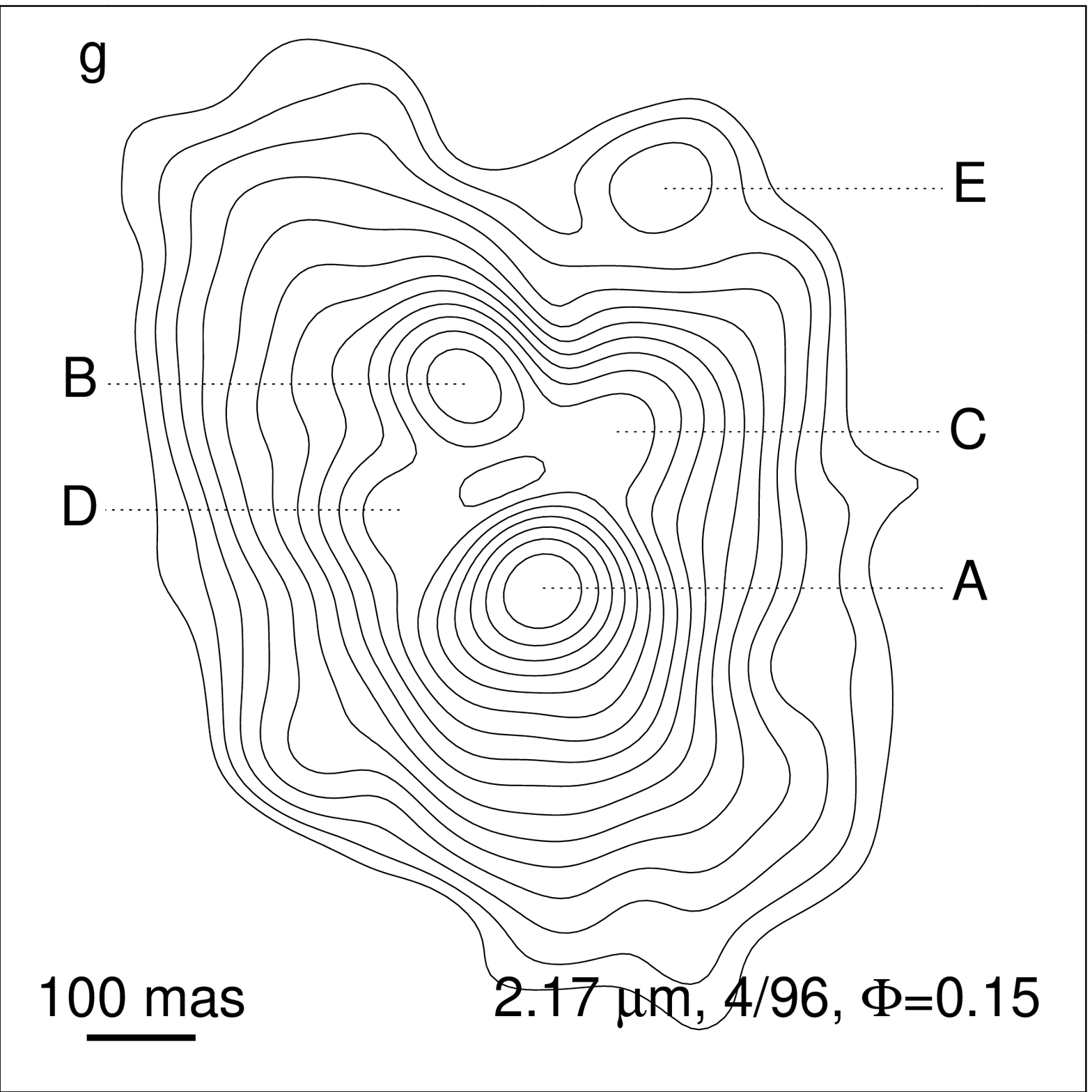}  }
    \put(73,146){ \epsfxsize=36mm\epsfbox{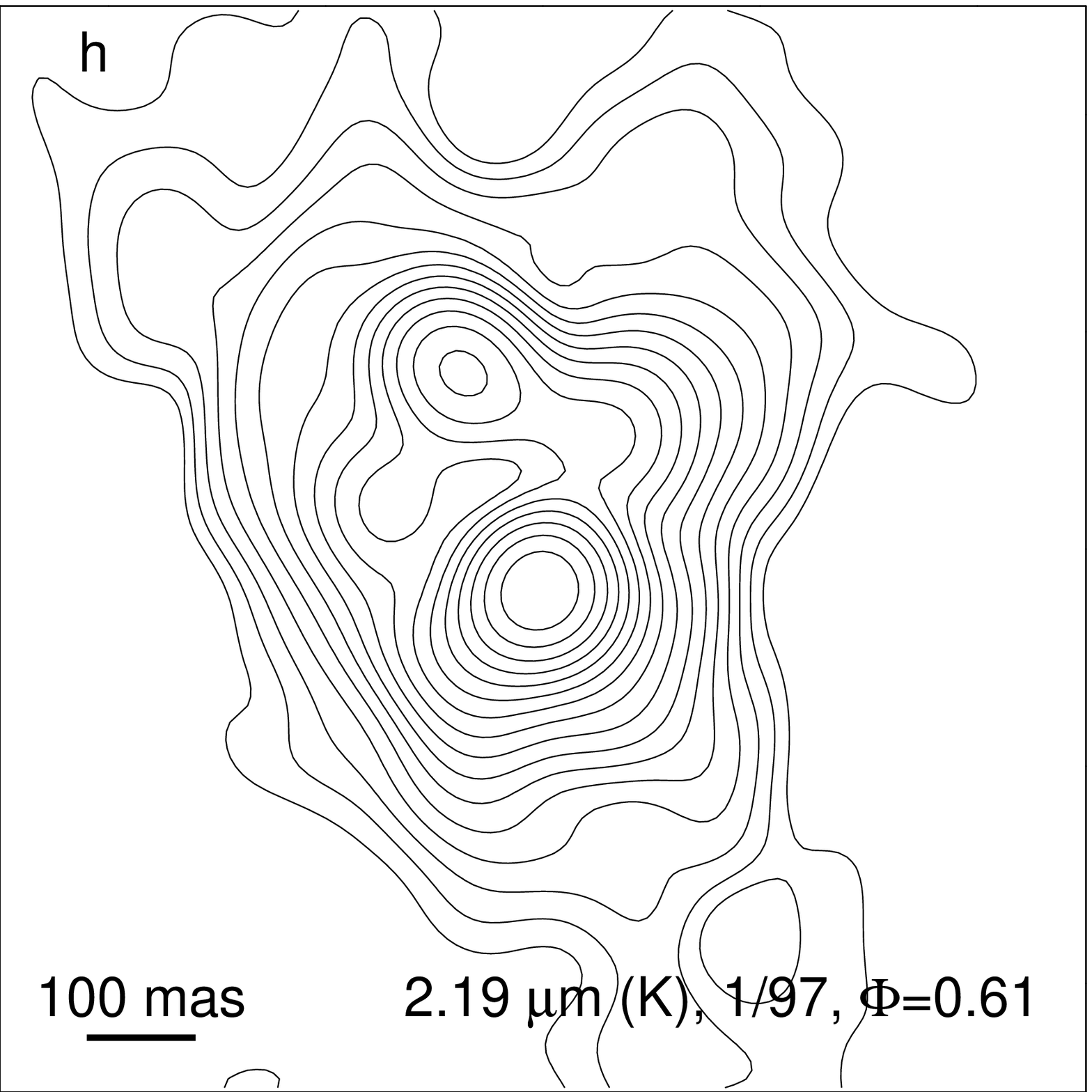}   }
    \put(73,73){  \epsfxsize=36mm\epsfbox{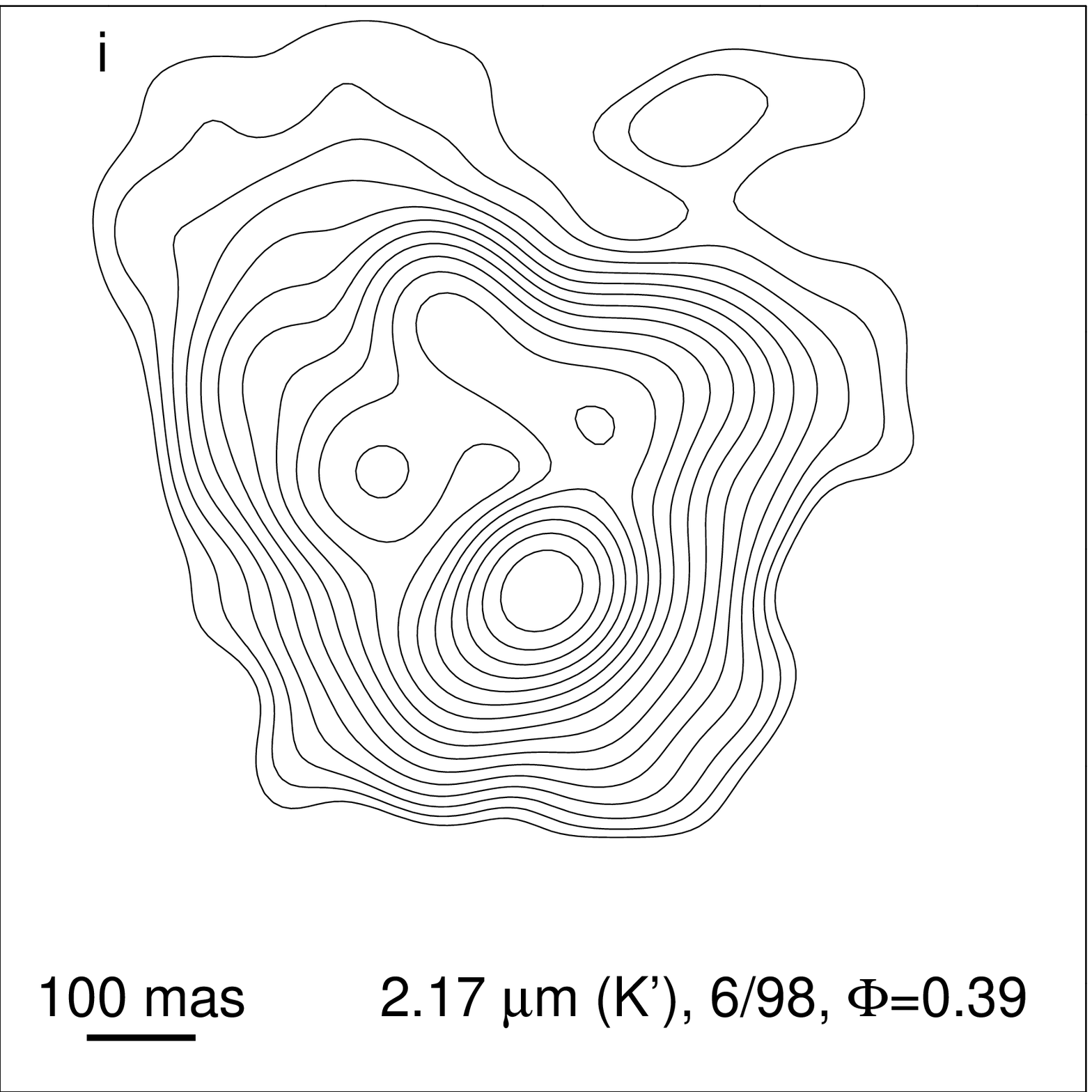}  }
    \put(73,0){   \epsfxsize=36mm\epsfbox{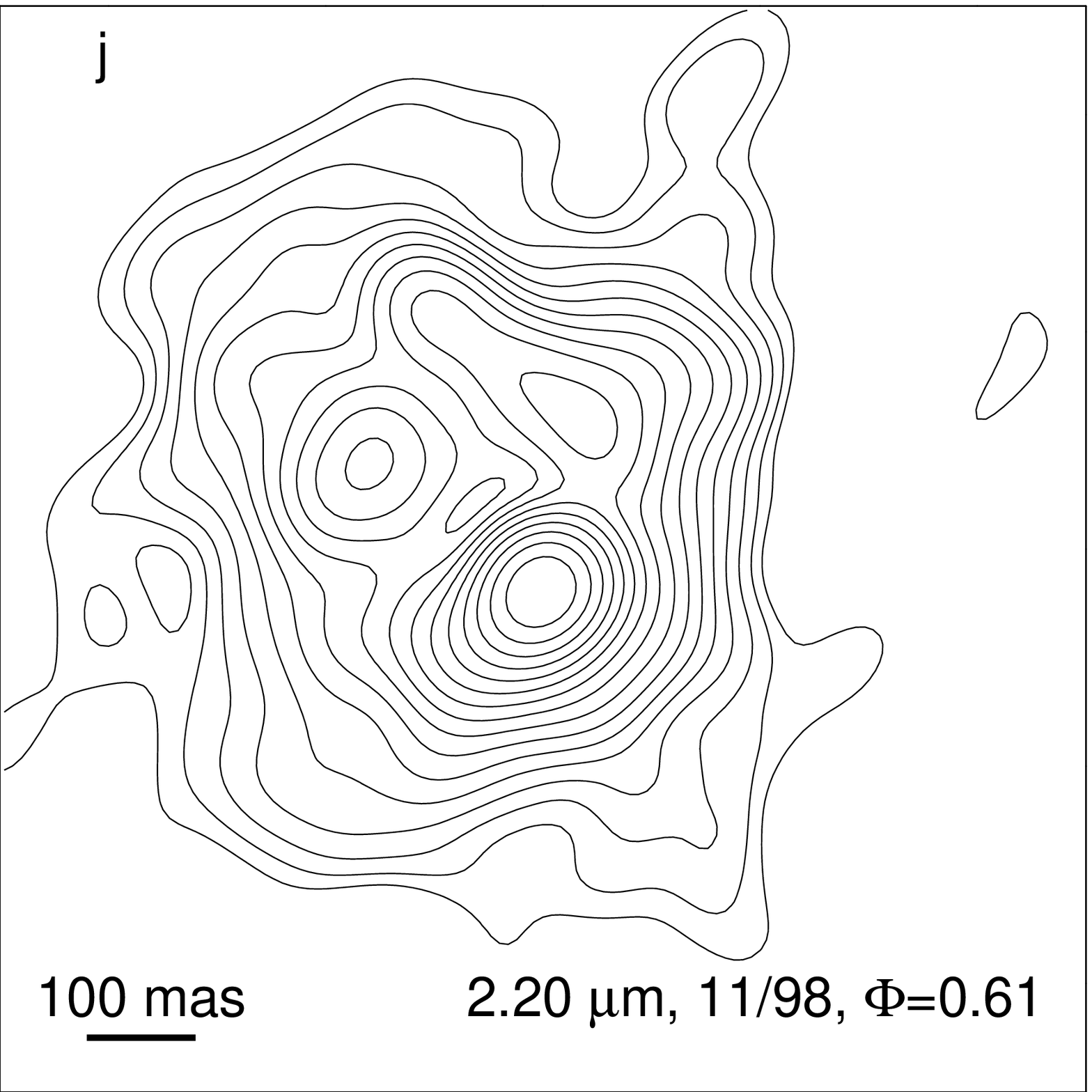} }
    \put(150,258){ \epsfxsize=45mm\epsfbox{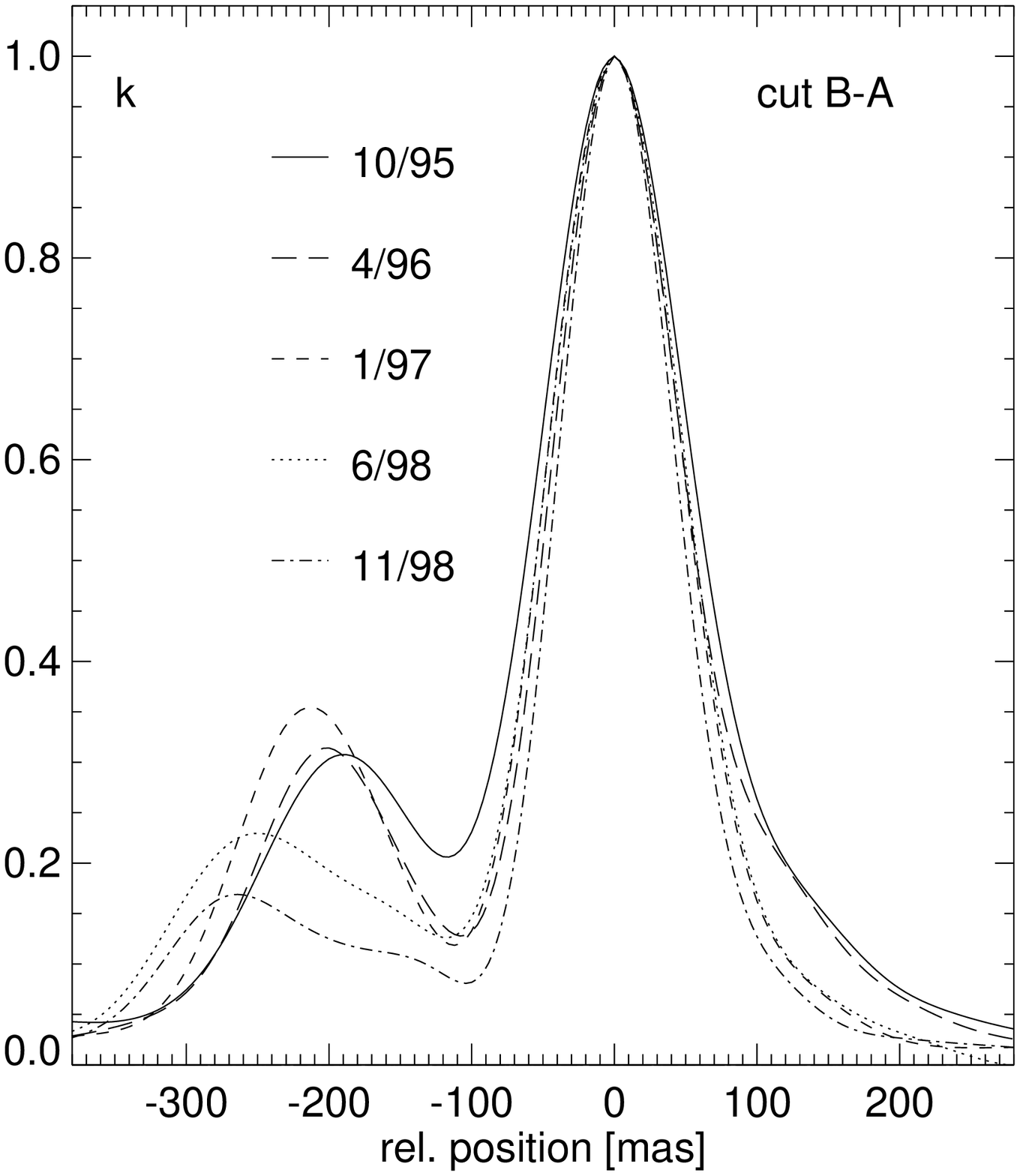} }
  \end{picture}
  \hfill
  \parbox[b]{40mm}{
  \caption{
    {\bf a} to {\bf e}:
    High--resolution bispectrum speckle interferometry images
    of IRC\,+10\,216.
    North is up and east to the left.  The figures represent a time
    series showing the evolution of the subarcsecond structure of
    IRC\,+10\,216 from 1995 (top) to 1998 (bottom).  In all
    figures the same gray level corresponds to the same relative
    intensity measured with respect to the peak.
    {\bf f} to {\bf j}:
    Same as a to e but as contour representation (contours at every 0.3~mag
    down to 4.8~mag relative to the respective peak).  The resolutions of
    the images are 92~mas (a, f), 82~mas (b, g), 87~mas (c, h), 87~mas
    (d, i), and 75~mas (e, j).
    In all figures 
    the filter wavelength, the epoch of the observation,
    and the photometric phase $\Phi$ are indicated.
    {\bf k}: Cuts through the images a to e along the axis from component A to
    B (position angle 20$^\circ$).
    \label{recall}
  }
  }
\end{figure}
\begin{figure}
  \setlength{\unitlength}{1mm}
  \begin{picture}(120,39)(0,0)
    \put(0,0){  \epsfxsize=39mm\epsfbox{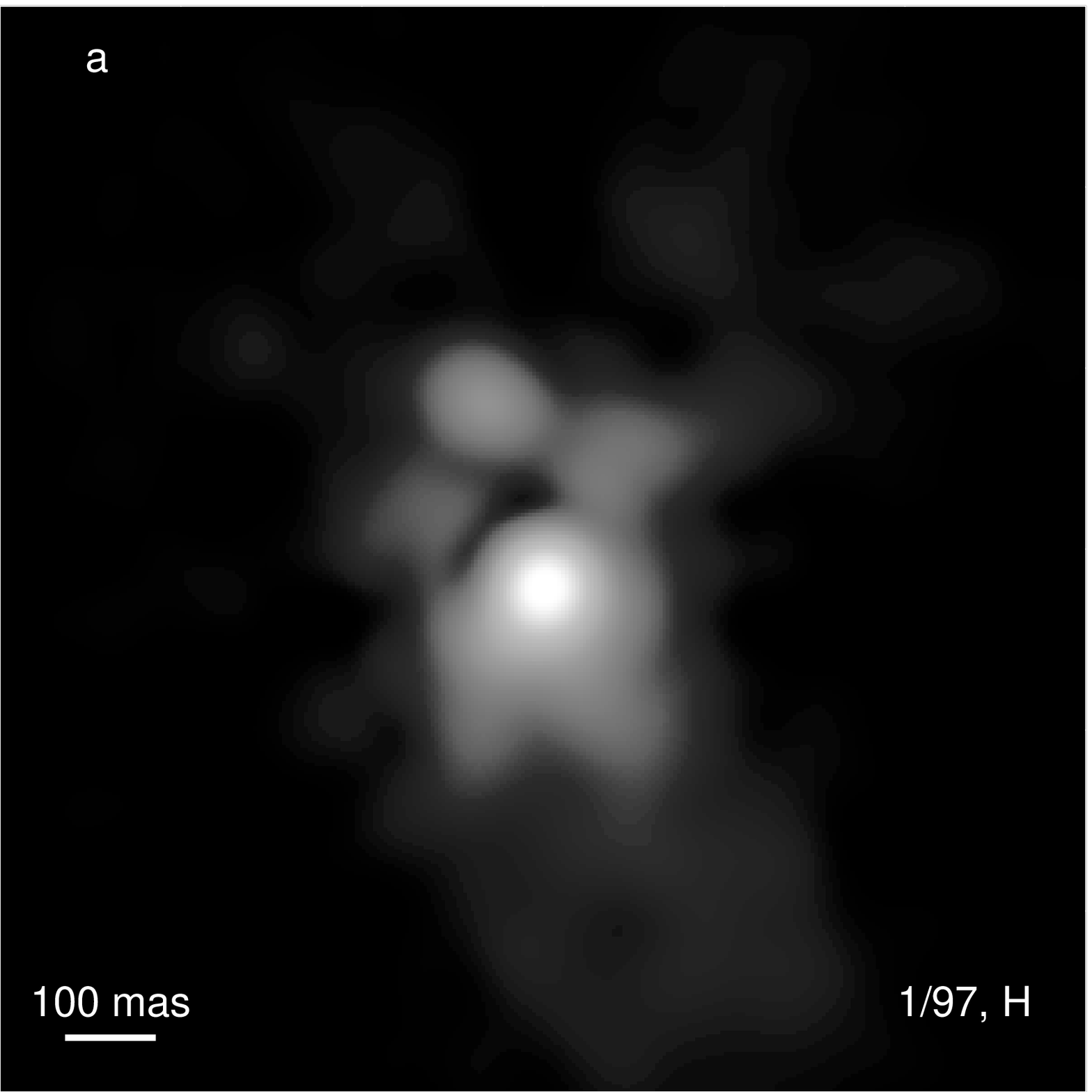}    }
    \put(40,0){ \epsfxsize=39mm\epsfbox{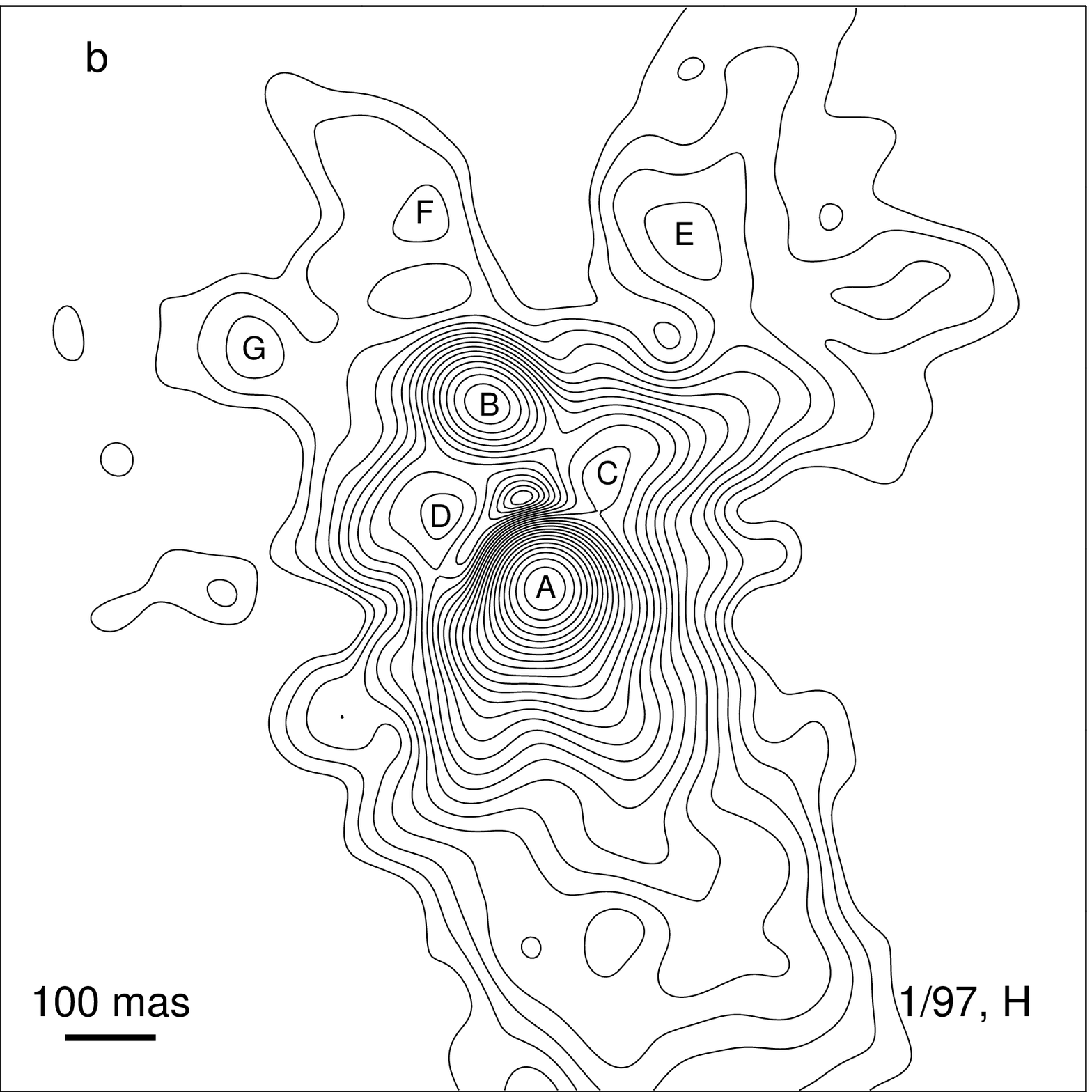} }
    \put(80,0){ \epsfxsize=39mm\epsfbox{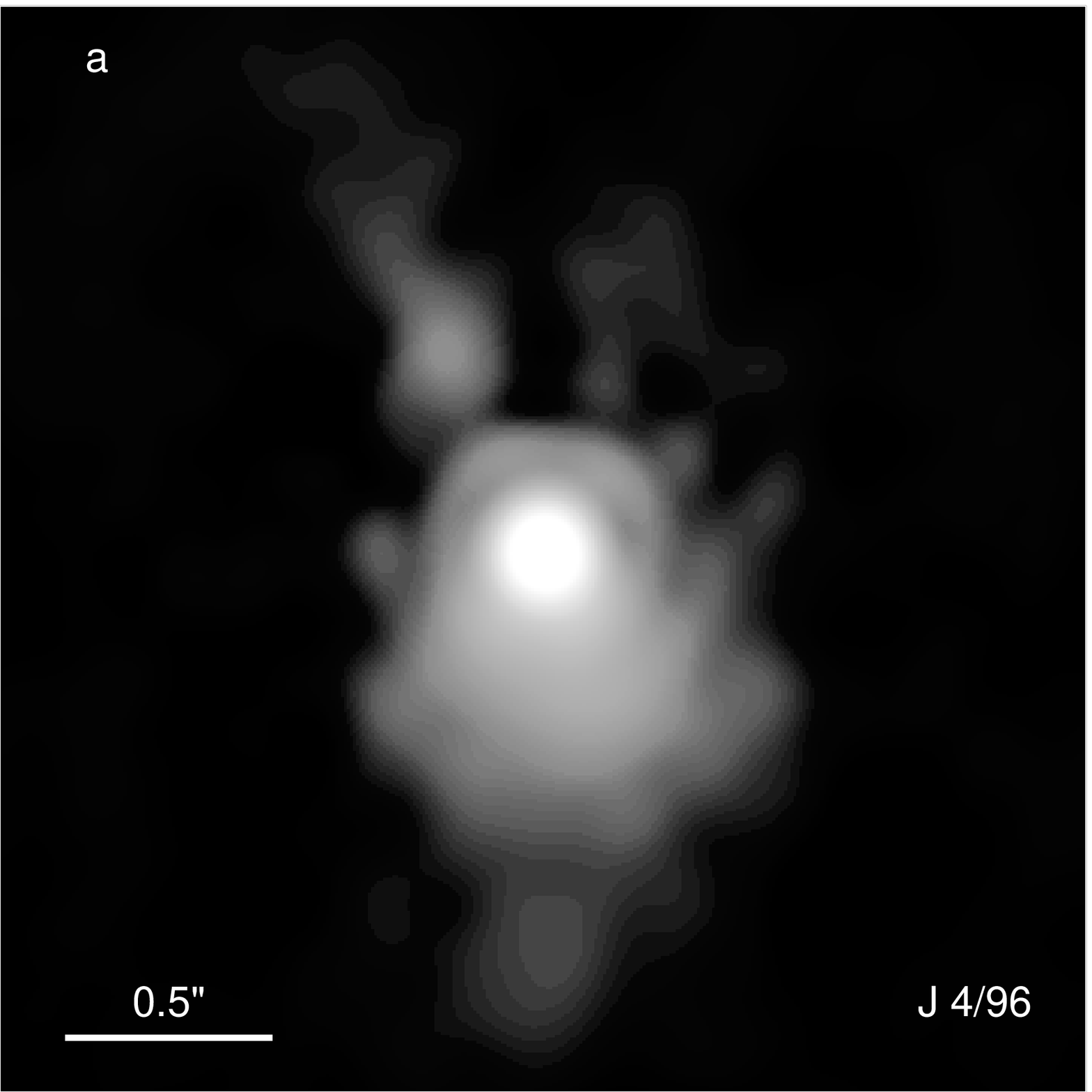}  }
  \end{picture}
  \caption{
    {\bf a}: 70~mas resolution bispectrum speckle interferometry image
    of IRC\,+10\,216 in the $H$--band.  North is up and east to
    the left.
    {\bf b}:
    Same as (a) as a contour image with denotations (A to G) of compact
    structures.  Contours are at every 0.2~mag down to 5.0~mag relative to the
    peak.
    {\bf c} $J$--band speckle reconstruction of IRC\,+10\,216 with 149~mas
	    resolution.
    \label{rech}
  }
\end{figure}

Figures~\ref{recall}-\ref{rech} show 
$J$-, $H$-, and $K$-band observations
of \inx{IRC\,+10\,216}\ (Osterbart et al.\ 2000).
The images were reconstructed from 6 m telescope speckle interferograms
using the bispectrum speckle interferometry method
(Weigelt 1977, Lohmann et al.\ 1983, Weigelt 1991).
The $H$ and $K$ images
with resolutions between 70~mas and 92~mas consist
of several compact components within a 0.2'' radius and a fainter
asymmetric nebula.
A series of $K$-band images from five epochs between Oct.\ 1995 and Nov.\ 1998
shows the dynamical evolution of the inner nebula 
(Fig.~\ref{recall}).
Denoting the  brightest four components with A to D
in order of decreasing brightness in the 1996 image,
the separation of the two brightest components A and B
increased from 191 mas in 1995 to 265 mas in 1998, i.e. by  $\sim35\%$,
corresponding to  a relative
velocity of  23~mas/yr or 14~km/s  within the plane of the sky at $d=130$\,pc.
Within these 3 years the rather faint components C and
D became brighter whereas component B faded.
Note that the observations cover more than one pulsational cycle.
Accordingly the apparent relative motions are not simply related
to stellar variability.
The general geometry
of the nebula seems to be bipolar, most prominently
present in the $J$-band image (Fig.~\ref{rech})
implying an asymmetric mass-loss.

The structures and changes in the inner nebula lead to the conclusion that
the central star is close to or at the position of component B.
Correspondingly, the initially brightest component A is the southern lobe of
a bipolar structure. 
Then, the star is strongly but not totally obscured
at $H$ and $K$.  Consistently, component B is very red in the $H-K$ color
while A and the northern $J$-band components are relatively blue.
This is supported by 1.1\,$\mu$m archival HST polarimetry data
(April 1997; see Osterbart et al.\ 2000).
The polarization pattern is predominantly centro-symmetric with its center
close to component B, whereas it shows strong polarization in the
northern arms and a still significant polarization at component A.
Finally, detailed two-dimensional radiative transfer calculations
(Men'shchikov et al.\ 2000; this volume) show that the observed
intensity ratio of A to B as well as the components' shapes clearly require
the star to be at B.
The inner nebula and the apparent motions seem to be rather symmetric around
this position and the  observed changes suggest an enhanced mass loss since
1997. A strongly variable mass loss operating on timescales of
several stellar pulsation periods has, in fact, been predicted by
dust-formation models of long-period carbon stars
(Winters et al.\ 1995; this volume).

\inx{IRC\,+10\,216}\ is without doubt in a very advanced stage of its
AGB evolution. The observed bipolarity of its dust shell even indicates that
it has possibly entered the phase of transformation into a protoplanetary
nebula. 
%
%
%
%
\section{H-deficient post-AGB stars}
The most likely scenarios invoked to explain the H-deficient WR central stars
are related to thermal pulses occuring either at the very end of the AGB
evolution or during the post-AGB stage.
During thermal pulses, recurrent instabilities of the He-burning shell,
the luminosity of the He shell increases rapidly for a
short time of 100\,yr to $10^{6}...\hspace*{0.5mm} 10^{8}$\,L$_{\odot}$.
The huge amount of energy produced forces the development of a pulse-driven
convection zone which mixes products of He burning, i.e.\ carbon and oxygen,
into the intershell region.
Because the hydrogen shell is pushed concomitantly into cooler domains,
hydrogen burning ceases temporarily allowing the envelope convection to proceed
downwards after the pulse,
to penetrate those intershell regions formerly enriched with
carbon (and oxygen), and to mix this material to the surface
(3$^{\rm rd}$ dredge up).
After the pulse hydrogen burning re-ignites and provides again the main
source of energy (see Bl\"ocker 1999 for a review).

\begin{figure}
\centering
\hspace*{0.4cm}
\epsfxsize=0.7\textwidth
\epsfbox{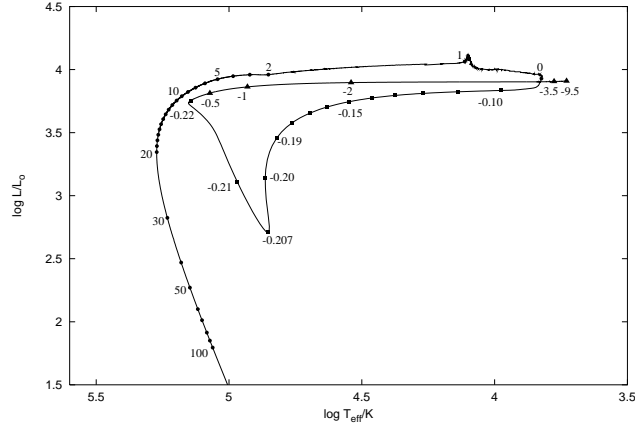}
\vspace*{-5mm}
\caption[track]{
Evolutionary track for a  0.625\,M$_{\odot}$ overshoot model suffering from an
LTP (Bl\"ocker 2000).
Symbols 
refer to time marks given in units of $10^{3}$\,yr.
Time is set to zero at minimum effective temperature after the flash.
The age -3500\,yr marks the begin of the central-star evolution and refers to
a pulsational period of 50\,d.
The preceding AGB evolution (Bl\"ocker 1995) did not include overshoot.
Since the intershell abundances of carbon and oxygen in overshoot AGB models
first increase and then level out after a series of thermal pulses, they
can be accounted for by calculating the pulse-driven convection zone with an
enhanced overshoot efficiency ($f$=0.064) leading to appropriate
intershell abundances of (He,C,O)=(45,40,13).
For the remaining evolution the standard efficiency ($f$=0.016;
Herwig et al.\ 1997) was used.
%
%
} \label{Fltptrack}
\end{figure}

\begin{figure}
\centering
\hspace*{0.4cm}
\epsfxsize=0.8\textwidth
\epsfbox[50 58 557 414]{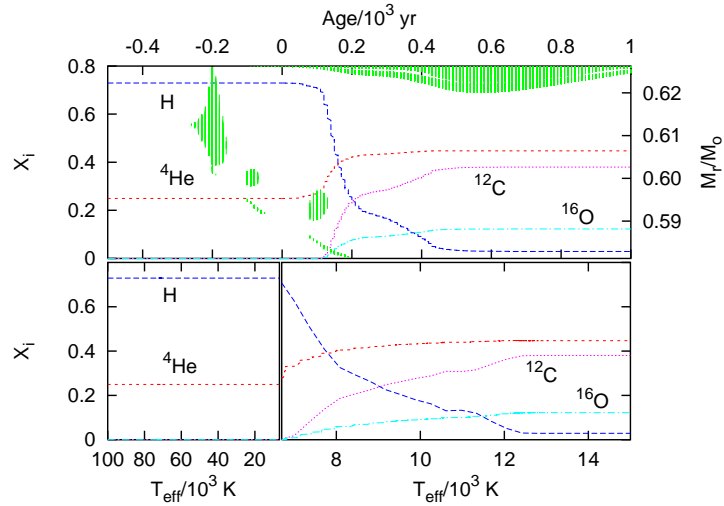}
\caption[heburn]{
{\bf Top}:
Evolution of the surface abundances of H, He, C, and O (left scale)
and extension of convective regions (right scale)
for the 0.625\,M$_{\odot}$ overshoot model in Fig.~\ref{Fltptrack}
during flash and  dredge up.
Shaded regions denote convective regions. Time is set to
zero at minimum effective temperature after the flash. 
{\bf Bottom}: Surface abundances of H, He, C, and O
as function of the effective temperature.
The line indicates the turn-around point at minimum effective temperature
after the flash corresponding to age zero in the above panel.
} \label{Fltpkipp}
\end{figure}

Herwig et al.\ (1997) showed that the consideration of diffusive overshoot
(Freytag et al.\ 1996) 
in all convective boundaries leads to important changes in AGB models, i.e. to
(i) efficient dredge-up and formation of low-mass carbon
stars;
(ii) formation of $^{13}$C as a neutron source to drive the $s$-process
in these stars; and
(iii) considerably changed intershell abundances.
The overshoot efficiency was calibrated
by the observed width of the main sequence.
The latter finding (iii) turned out to be a key ingredient for the modelling of
WR central stars. Overshoot leads
to an enlargement of the pulse-driven convection zone and
to enhanced mixing of core material from deep layers below the He shell
to the intershell zone
resulting in intershell abundances (mass fractions) of (He,C,O)=(40,40,16)
instead of (70,25,2) as in non-overshoot sequences.
These modified intershell abundances are already close to the 
observed surface abundances of Wolf-Rayet central stars.
Finally, in contrast to standard evolutionary calculations, 
overshoot models do show dredge up for very low envelope masses 
and efficient dredge up was found even during the post-AGB 
stage (Bl\"ocker 2000, Herwig 2000) leading to the mixture of the intershell
abundances to the surface and to the dilution of hydrogen. 
Three thermal pulse scenarios for Wolf-Rayet central stars can now
be distinguished:  \\
{\bf AGB Final Thermal Pulse (AFTP)}, 
occurring immediately before the star moves off the AGB.
In this case the envelope mass is already very small
($\sim 10^{-2} $M$_{\odot}$). During dredge-up
a substantial fraction of the intershell region is mixed with the
tiny envelope leading to the dilution of hydrogen and enrichment
with carbon and oxygen. The resulting surface abundances
depend on the actual envelope mass at which the AFTP occurs.
For instance, 
Herwig (2000) found for M$_{\rm env}= 4 \cdot 10^{-3} $M$_{\odot}$ 
(H,He,C,O)=(17,33,32,15) 
after an AFTP.
The AFTP leads to a relatively high hydrogen abundance ($\ga 15\%$)
and predicts small kinematical ages for the PNe of WR central stars which 
emerge here directly from the AGB. \\
{\bf Late Thermal Pulse (LTP)},
occurring when the model evolves
with roughly constant luminosity from the AGB towards the white dwarf
domain. This kind of thermal pulse is similar to the one experienced by
AGB stars but the envelope mass is even smaller than for the AFTP
($\sim 10^{-4} $M$_{\odot}$).
Fig.~\ref{Fltptrack} shows the evolutionary track of a 0.625\,M$_{\odot}$
LTP model with diffusive overshoot (Bl\"ocker 2000).
%
After the flash
the intershell abundances amounted to (He,C,O)=(45,40,13) and
the model evolves
towards the AGB domain. At minimum effective temperature ($\approx 6700$\,K)
dredge up sets in 
and continues until the star has reheated to $\approx 12000$\,K.
Hydrogen is diluted to 3\% and the final surface abundances of He, C and O 
are close to those of the intershell region, viz.\ (45,38,12). Extension of 
convective regions and abundances are illustrated in Fig.~\ref{Fltpkipp} as 
function of age and effective temperature,
resp. The kinematical age of the PN amount to a few thousand years . 
\\
{\bf Very  Late Thermal Pulse (VLTP)},
occurring when the model is already on the  white dwarf cooling track,
i.e.\ after the cessation of  H burning. 
Then, the pulse-driven convection zone can reach and penetrate the H-rich
envelope and protons are ingested into the hot, carbon-rich intershell 
region 
raising a H flash (Iben \& McDonald 1995).
The energy released by this flash leads to a splitting of the convection 
into an upper zone powered by H burning and a lower one powered by He burning. 
The upper convection zone is, however, short-lived because the available 
hydrogen in the envelope is quickly consumed. Finally, the star becomes 
hydrogen-free and exposes its intershell abundances at the surface. 
Herwig et al.\ (1999) found for a 0.604\,M$_{\odot}$ overshoot model
surface abundances of (He,C,O)=(38,36,17). 
The kinematical age of the PN is relatively high
since the star has first to fade along the cooling branch down
to a few 100\,L$_{\odot}$ before the flash. 
For 0.6 M$_{\odot}$ one obtains typically $t \ga  20000$\,yr. \\[0.5ex]
%
All scenarios lead to hydrogen-deficient post-AGB stars with 
carbon and oxygen abundances as observed for Wolf-Rayet central stars.
The variety of observations requires most likely all of these scenarios.
Many objects have only very low hydrogen abundances, if any, favoring the 
LTP and VLTP. On the other hand, several Wolf-Rayet central stars are 
surrounded by young  planetary nebulae (Tylenda 1996) and circumstellar shells 
(Waters et al.\ 1998) strengthening the AFTP and LTP. 
Within the current models roughly 20 to 25\% of stars moving off the AGB 
can be expected to become hydrogen-deficient.

{}
\end{document}